\pdfoutput=1

\PassOptionsToPackage{
  ocgcolorlinks,%
  breaklinks,%
  hidelinks,
  pdfauthor={Inge Madshaven et al.},
  pdftitle={
    Cerman:
    Software for simulating streamer propagation
    in dielectric liquids
    based on
    the Townsend--Meek criterion},
}{hyperref}

\documentclass[
    times,          
    final,          %
    english,        
    ]{elsarticle}   

\usepackage{framed}
\usepackage{multirow}

\usepackage{fancyhdr}

\usepackage{etoolbox}        
\usepackage{amsmath}         
\usepackage[utf8]{inputenc}  

\usepackage{lmodern}         
\usepackage{bm}
\usepackage{xcolor}
\usepackage{footmisc}     
\usepackage{textcomp}     
\usepackage{microtype}

\usepackage{graphics}     
\usepackage{booktabs}     
\usepackage{tabularx}
\usepackage{url}          
\usepackage{tikz}         

\usepackage{float}     
\usepackage{listings}  

\usepackage[detect-all,per-mode=symbol]{siunitx}

\usepackage{natbib}

\usepackage{hyperref}

\usepackage{cleveref}
\crefname{equation}{}{}
\Crefname{equation}{Equation}{Equations}
\crefname{figure}{figure}{figures}
\Crefname{figure}{Figure}{Figures}
\crefname{table}{table}{tables}
\Crefname{table}{Table}{Tables}
\crefname{appendix}{}{}  
\Crefname{appendix}{}{}
\crefname{lstfloat}{listing}{listings}
\Crefname{lstfloat}{Listing}{Listings}
\crefname{algfloat}{algorithm}{algorithms}
\Crefname{algfloat}{Algorithm}{Algorithms}

\renewcommand{\vec}[1]{\bm{#1}}
\let\oldhat\hat
\renewcommand{\hat}[1]{\oldhat{\bm{#1}}}

\colorlet{punct}{red!60!black}
\definecolor{background}{HTML}{EEEEEE}
\definecolor{delim}{RGB}{20,105,176}
\colorlet{numb}{magenta!60!black}

\floatstyle{plaintop}  
\newfloat{lstfloat}{tbph}{lop}
\floatname{lstfloat}{Listing}

\floatstyle{plaintop}  
\newfloat{algfloat}{tbph}{lop}
\floatname{algfloat}{Algorithm}

\definecolor{blue(ncs)}{rgb}{0.0, 0.53, 0.74}
\definecolor{jade}{rgb}{0.0, 0.66, 0.42}
\definecolor{maize}{rgb}{0.98, 0.93, 0.37}
\definecolor{ao(english)}{rgb}{0.0, 0.5, 0.0}
\definecolor{beigg1}{HTML}{FFFFF3}
\definecolor{beigg2}{HTML}{EBEBC7}
\definecolor{blue(pigment)}{rgb}{0.2, 0.2, 0.6}
\definecolor{githubgray}{HTML}{EFF0F1}
\definecolor{github-bg-gray}{HTML}{f6f8fa}
\definecolor{github-bg-light-gray}{HTML}{fafbfc}

\lstdefinelanguage{fish}  
{
  basicstyle=\ttfamily\bfseries,
  showstringspaces    = false,
  keywords            = {cerman, pip},
  keywordstyle        = \ttfamily\bfseries,
  alsoletter          = 0123456789.+-,
  morestring          = [s]{"}{"},
  morestring          = [s][\color{blue(ncs)}]{<}{>},
  morecomment         = [l][\color{ao(english)}]{\#},
  xleftmargin=0em, xrightmargin=0em,
}

\lstdefinelanguage{fish-gray}  
{
  backgroundcolor=\color{githubgray},
  basicstyle=\ttfamily,
  showstringspaces    = false,
  keywords            = {cerman, pip},
  keywordstyle        = \ttfamily\bfseries,
  alsoletter          = 0123456789.+-,
  morestring          = [s]{"}{"},
  morestring          = [s]{<}{>},
  morecomment         = [l][\color{github-bg-gray}]{\#},
  xleftmargin=0em, xrightmargin=0em,
}

\makeatletter
\newif\ifcolonfoundonthisline
\newcommand\JSONnumbervaluestyle{\color{blue(ncs)}\bfseries}
\newcommand\JSONstringvaluestyle{\color{ao(english)}\bfseries}

\lstdefinelanguage{json}
{
  frame=single,
  framerule=1pt,
  rulecolor=\color{beigg2},
  backgroundcolor=\color{beigg1},
  basicstyle=\ttfamily\small,
  showstringspaces    = false,
  keywords            = {false,true,null},
  keywordstyle        = \ttfamily\bfseries,
  alsoletter          = 0123456789.+-,
  morestring          = [s]{"}{"},
  stringstyle         = \ifcolonfoundonthisline\JSONstringvaluestyle\else\fi,
  MoreSelectCharTable =%
    \lst@DefSaveDef{`:}\colon@json{\processColon@json}{},
  xleftmargin=0em, xrightmargin=0em,
}

\newcommand\processColon@json{%
  \colon@json%
  \ifnum\lst@mode=\lst@Pmode%
    \global\colonfoundonthislinetrue%
  \else\fi
}

\lst@AddToHook{Output}{%
    \ifnum\lst@mode=\lst@Pmode%
      \def\lst@thestyle{\JSONnumbervaluestyle}%
    \else\fi
  \lsthk@DetectKeywords%
}

\lst@AddToHook{EOL}%
  {\global\colonfoundonthislinefalse}

\makeatother

\makeatletter
\begingroup
\catcode`_=\active
\catcode`^=\active
\protected\gdef_{\@ifnextchar'{\subtextrm}{\@ifnextchar"{\subtextsc}{\sb}}}
\protected\gdef^{\@ifnextchar'{\suptextrm}{\@ifnextchar"{\suptextsc}{\sp}}}
\endgroup
\def\subtextrm'#1'{\sb{\textrm{#1}}}
\def\suptextrm'#1'{\sp{\textrm{#1}}}
\def\subtextsc"#1"{\sb{\textsc{#1}}}
\def\suptextsc"#1"{\sp{\textsc{#1}}}
\AtBeginDocument{\catcode`_=12 \mathcode`_="8000}
\AtBeginDocument{\catcode`^=12 \mathcode`^="8000}
\makeatother

\newcommand{\abs}[1]{{\lvert#1\rvert}}

\newcommand{\mum}{\si{\micro \meter}}
\newcommand{\mus}{\si{\micro \second}}

\newcommand{\json}{{JSON}}
\newcommand{\roi}{{ROI}}
\newcommand{\ROI}{{ROI}}
\newcommand{\cfd}{CFD}

\newcommand{\ilcode}[1]{{%
  \setlength{\fboxsep}{1 pt}%
  \colorbox{githubgray}%
  {%
    \vphantom{gt}%
    \texttt{\hspace*{0pt}#1\hspace*{0pt}}}%
  }}

\begin{document}


\begin{frontmatter}

\begin{abstract}%
%
We present a software to simulate
the propagation of positive streamers in dielectric liquids.
Such liquids are commonly used for electric insulation of high-power equipment.
We simulate electrical breakdown in a needle--plane geometry,
where the needle and the extremities of the streamer are modeled by hyperboloids,
which are used to calculate the electric field in the liquid.
If the field is sufficiently high,
electrons released from anions in the liquid can turn into electron avalanches,
and the streamer propagates if an avalanche meets the Townsend--Meek criterion.
The software is written entirely in Python and released under an MIT license.
We also present a set of model simulations demonstrating the capability
and
versatility of the software.
%
\end{abstract}



\title{%
  Cerman:~
  Software for simulating streamer propagation
  in dielectric liquids
  based on\\
  the Townsend--Meek criterion
  }


\newcommand{\snm}[1]{#1}

\author[ntnu]{I.~\snm{Madshaven}}

\author[sintef]{O. L. \snm{Hestad}}

\author[ntnu]{P.-O.~\snm{Åstrand}\corref{cor1}}
\cortext[cor1]{Corresponding author: \texttt{per-olof.aastrand@ntnu.no}}

\address[ntnu]{%
  Department of Chemistry,
  NTNU --\\ Norwegian University of Science and Technology,
  7491 Trondheim, Norway%
  }
\address[sintef]{%
  SINTEF Energy Research,
  7465 Trondheim, Norway%
  }

\end{frontmatter}

\pagestyle{fancy}
\renewcommand{\headrulewidth}{0pt}  
\lhead{\emph{Cerman}}  
\rhead{I. Madshaven \textit{et al.}}  


\section{Introduction}{\label{sec:introduction}

\subsection{Streamers in liquids}

Dielectric liquids, specifically transformer oils, are used as electric insulation
in high-power equipment such as power transformers~%
\cite{Wedin2014}.
Equipment failure is always a possibility,
and in a world with ever-growing need for energy,
there is a continuous effort to make equipment better, cheaper,
more compact,
and
more environmentally friendly.
To prevent equipment failure due to electrical discharges,
new insulating liquids as well as additives are tested,
experiments are carried out
to better understand the physical nature of the phenomena,
and
simulations are performed
to test the validity of predictive models~%
\cite{Lesaint2016cxmf,Sun2016dbt2}.

Since electrical discharge events are rare
at operating conditions,
model experiments are designed to induce discharge in the liquid.
In one such model experiment,
a needle electrode is placed opposing a planar electrode,
where the needle--plane gap is insulated by a liquid~%
\cite{Lesaint2016cxmf}.
If high voltage is applied,
resulting in a sufficiently strong electric field close to the needle,
the liquid will lose its insulating properties
and begin to conduct electricity,
and subsequent (partial) discharges
from the needle into the liquid can occur.
The charge transported into the liquid
can increase the electric field and lead to
partial discharges in new regions
in a self-induced manner.
Shadowgraphic images
(an imaging technique exploiting differences in permittivity)
reveal that
a gaseous channel, a ``streamer'',
is formed
and
how it branches as it propagates through the liquid~%
\cite{Farazmand1961bhcrhp}.
If a streamer bridges the gap between two electrodes,
an electric discharge can follow,
possibly destroying the affected equipment.

Streamers are commonly classified by their polarity
and speed of propagation
from the slow 1st mode to the fast 4th mode,
ranging from below \SI{0.1}{km/s} to well above \SI{100}{km/s}~%
\cite{Lesaint2016cxmf}.
Streamers with negative polarity
typically have a lower inception voltage
than streamers with positive polarity (positive streamers),
however,
positive streamers typically lead to breakdowns
at lower voltage than negative streamers,
and as such,
research is mainly concerned with positive streamers.
The streamer phenomenon involves
processes covering several length and time scales.
Speed and branching is studied
in gaps of different sizes (\si{mm}--\si{m}),
while many of the interesting processes,
such as
field ionization,
high-field conduction,
electro-hydrodynamic movement,
bubble nucleation,
cavitation,
electron avalanches,
photoionization,
occur on a \mum-scale~%
\cite{Lesaint2016cxmf,Sun2016dbt2}.
A streamer usually stops or leads to a breakdown
on a \mus-scale ($\si{km/s} = \si{mm/\micro s}$),
whereas processes such as
recombination of electrons and anions
can occur within picoseconds.  
Consequently,
experimentation as well as simulation is challenging.

\subsection{Modeling and simulations}

While sophisticated equipment is required for experiments,
simulations often investigate
the effect of given processes through relatively simple models.
The fractal nature of the streamer structure can be simulated
by considering a lattice where each point is either
part of
an electrode, the liquid, or the streamer~%
\cite{Niemeyer1984d35qr4}.
Here, the streamer expands to new lattice points when some criterion,
such as electric field strength,
is obtained.
Through kinetic Monte Carlo methods,
the stochastic nature
and
the physical time
can also be studied in such simulations~%
\cite{Biller1993dctxz7}.
By considering the streamer as
an electrical network of resistors and capacitors,
the charge and conduction can be studied,
without necessarily confining the points of the network to a grid~%
\cite{Fofana1998bm4sd5}.
Models where the streamer consists of a set of discrete points
are simplistic but also efficient.
Conversely,
with a higher demand for computational power,
computational fluid dynamic (\cfd) methods
can be applied to solve
the equations for generation and transport of charged particles
(the flow of natural particles is often ignored)
during a streamer discharge~%
\cite{Qian2006c75hff,Simonovic2019dcjj},
while
the stochasticity and branching of streamers can be introduced
by adding impurities~%
\cite{Jadidian2014f55gj5}.
Such \cfd-calculations often ignore
the phase change from liquid to gas as well
and
are confined to a small region
because of the computational complexity.
However,
for simplified, single-channel streamers,
both the phase change
and
the processes in the channel
can be simulated~%
\cite{Naidis2015cxmj,Naidis2016cxmg}.
Code used for simulation of streamers in liquids is rarely published,
in fact, we found just a single example~\cite{Fowler2003ffwcfd}.

\subsection{Avalanche model}\label{sec:avamod}

\begin{figure}[t!]
    \centering
    \includegraphics[width=0.8\linewidth]{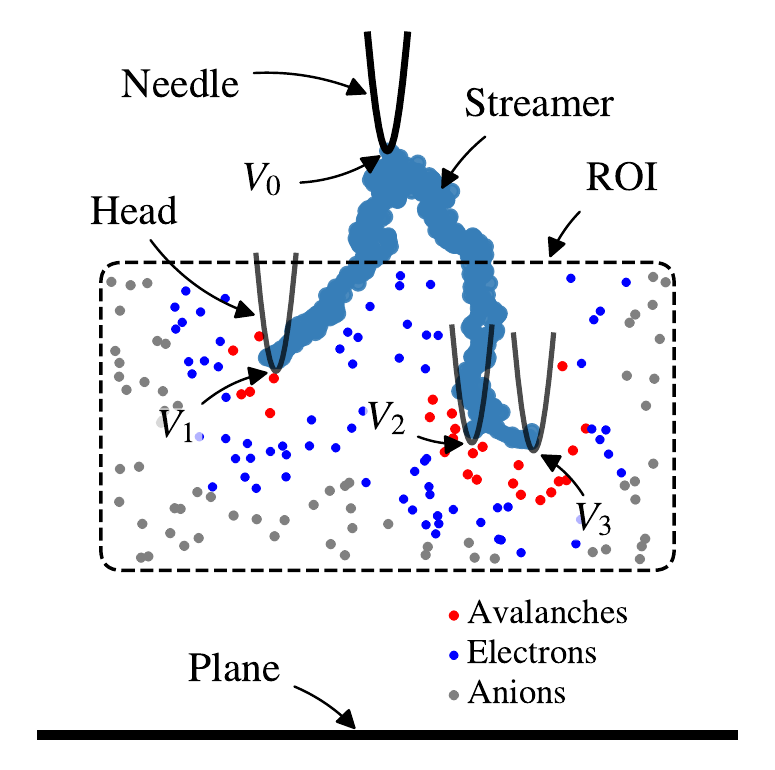}
    \caption{
        Illustration of the main components in the simulation model.
        The needle electrode and the streamer heads are hyperboloids,
        each with a potential $V_i$.
        A region of interest (\roi) is used to limit the computational effort
        to a region surrounding the active part of streamer.
        The \roi\ controls the position of the ``seeds'',
        which are classified as anions, electrons, or avalanches,
        depending on the electric field strength at their position.
        The ``shadowgraphic'' image of the streamer is created by
        plotting all former positions of streamer heads.
    }
    \label{fig:seeds_hyper}
\end{figure}

We have previously described our streamer model for positive streamers
where the propagation is based on an electron avalanche mechanism~%
\cite{Madshaven2018cxjf}.
The model has been extended to account
for conductance in the streamer channel
and capacitance between the streamer and the planar electrode~%
\cite{Madshaven2019c933},
as well as photoionization in front of the streamer~%
\cite{Madshaven2020dg8m},
the latter as a mechanism for the transition from slow to fast propagation.

Streamer propagation is simulated in a setup resembling model experiments,
a needle--plane gap filled with a model liquid,
see \cref{fig:seeds_hyper} for details.
The needle and streamer give rise to an electric field,
affecting charged particles in the liquid.
A number of anions,
``seeds'' for electron avalanches,
is modeled within a volume surrounding the streamer,
a ``region of interest'' (\roi).
Electrons released from the anions
can create electron avalanches,
and the streamer propagates
when an avalanche meets the Townsend--Meek criterion,
i.e.\ exceeds a critical number of electrons~%
\cite{Madshaven2018cxjf}.
The needle and the streamer heads (the extremities of the streamer)
are modeled as hyperboloids,
which simplifies calculating the electrical field
since the Laplacian is analytic in prolate spheroid coordinates~%
\cite{Coelho1971}.
The electric field and potential is calculated
by considering electrostatic shielding~%
\cite{Madshaven2018cxjf},
as well as
the conductance in the channel
and
the capacitance towards the planar electrode~%
\cite{Madshaven2019c933}.
The streamer undergoes a transition into a fast propagation mode
when radiation from the streamer head
can ionize molecules directly in front of the streamer~%
\cite{Madshaven2020dg8m}.
More details on the model is given in \cref{sec:implementation}.

The main output of the simulations include
the propagation speed,
the streamer shape (branching),
and
propagation distance.
In addition,
properties such as
the initiation time,
the potential of individual streamer heads,
electric breakdown within the streamer channel,
and
avalanche growth,
can also be investigated.
Simulations show how various parameters affect the results,
where the gap size, applied voltage, and type of liquid
are important parameters for a simulation.
Furthermore,
other parameters such as
the size of a streamer head,
the conductivity of the streamer channel,
properties of additives,
and
avalanche growth parameters
can be varied to
validate whether the underlying physical models are reasonable.

\subsection{Scope}

The present work describes
the use, functionality and implementation
of \emph{Cerman}~\cite{streamergit},
a software to do simulations with our model~%
\cite{Madshaven2018cxjf,Madshaven2019c933,Madshaven2020dg8m},
with the purpose to make the software publicly available.
\Cref{sec:simulation}
demonstrates how to
set up,
simulate,
and
evaluate results
of a relevant problem.
Further details on the model and its implementation
are given in \cref{sec:implementation},
whereas \cref{sec:functionallity}
outlines the current functionality
and some prospects for the future.
A summary is then given in \cref{sec:conclusion}.
Furthermore,
details on the algorithm is included in \cref{sec:algorithm},
simulation parameters in \cref{sec:parameters},
and simulation example input files in \cref{sec:examplefiles}.

}
%


\section{Simulation -- using the software}{\label{sec:simulation}

\begin{lstfloat}
    \caption{
        Example of \json-input file, \protect\ilcode{cmsim.json},
        defining a simulation series
        with several values for
        the needle voltage $V_0$
        and
        the threshold for breakdown in the streamer channel $E_'bd'$,
        both with and without photoionization enabled.
        Furthermore,
        each permutation is to be carried out 10 times
        with different initial seed positions.
        Note, by setting \protect\ilcode{alphakind} to \protect\ilcode{A1991},
        $\alpha$ is calculated by \protect\cref{eq:alpha_atraz}.
        See \protect\cref{sec:parameters} for a description of the parameters.
        }
    \vspace*{1.5 ex}
\begin{lstlisting}[language=json]
{
    "gap_size": 0.010,
    "needle_radius": 6e-06,
    "needle_voltage": "linspace(60e3, 150e3, 10)",
    "liquid_name": "cyclohexane",
    "alphakind": "A1991",
    "liquid_Ealpha": 0.65e09,
    "liquid_IP": 9,
    "streamer_head_radius": 3e-06,
    "streamer_U_grad": 0e+06,
    "streamer_d_merge": 12.5e-6,
    "streamer_scale_tvl": 0.1,
    "streamer_photo_enabled": [false, true],
    "streamer_photo_efield": 3.1e9,
    "streamer_photo_speed": 20e3,
    "rc_tau0": 1e-6,
    "rc_breakdown": [0e6, 4e6, 8e6, 16e6, 64e6],
    "random_seed": 1,
    "simulation_runs": 10,
    "save_specs_enabled": {
        "stat": true,
        "gp5": true
    }
}
\end{lstlisting}
    \label{lst:sim_series_example}
\end{lstfloat}

\subsection{Software overview}
The software name \emph{Cerman} is an abbreviation of \emph{ceraunomancy},
which means to control lightning
or
to use lightning to gain information.
The implementation is done in Python,
an open-source, interpreted, high-level, dynamic programming language%
~\cite{pythonorg},
and the software is available on
\href{https://github.com/madshaven/cerman}{GitHub}~\cite{streamergit}
under an \href{https://en.wikipedia.org/wiki/MIT_License}{MIT license}.
The software is script-based,
and controlled through
the command \ilcode{cerman},
which is used for
creation of input files,
running simulations,
and evaluating the results.
When a simulation is started,
the simulation parameters are loaded from an input file,
and the classes for the various functions are initiated.
See \cref{sec:parameters} for a summary of simulation parameters.
The simulation itself is essentially a loop where
seeds in the liquid are moved and the streamer structure is updated
until the streamer stops or leads to a breakdown.
The algorithm is detailed in \cref{sec:algorithm}.

\subsection{Getting started}

Download Cerman from
\href{https://github.com/madshaven/cerman}{GitHub}~\cite{streamergit}
and install it by running
\begin{lstlisting}[language=fish]
    pip install .
\end{lstlisting}
from the downloaded folder.
This installs the python package
and
the script \ilcode{cerman}.
Python 3.6 or above is required,
as well as the packages
\ilcode{numpy}~\cite{Oliphant2007},
\ilcode{scipy}~\cite{Virtanen2020ggj45f},
\ilcode{matplotlib},
\ilcode{simplejson},
and
\ilcode{statsmodels}.
The dependencies are automatically installed by \ilcode{pip}.
The software has been developed in OSX
and has been tested on Linux as well.

\subsection{Create simulation input}\label{input}

\begin{figure}
    \centering
    \includegraphics[width=0.8\linewidth]{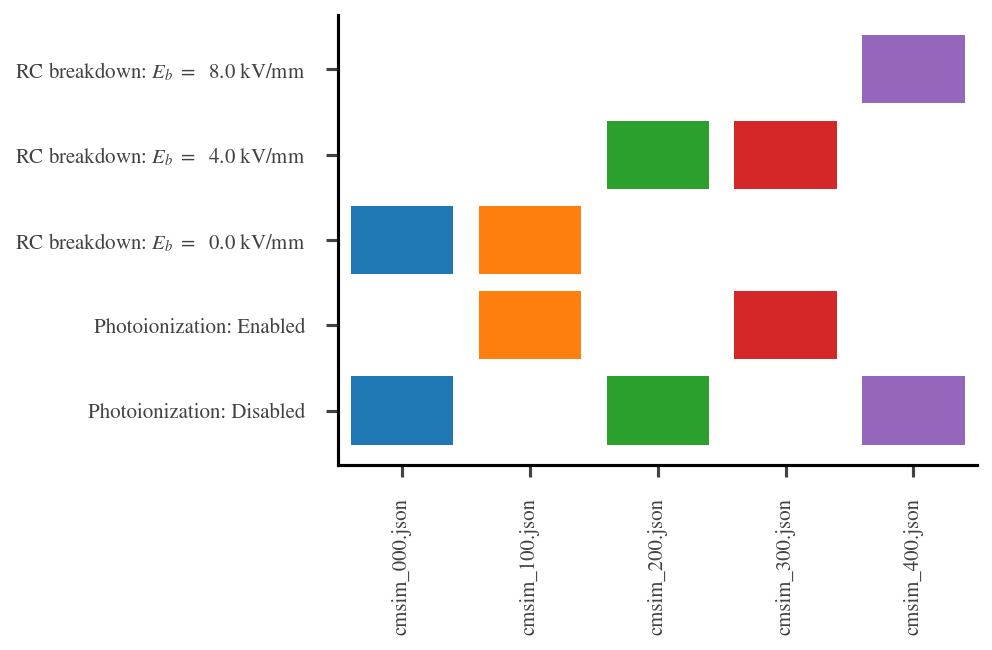}
    \caption{
        Visualization of the difference
        in parameter values between a selection of input files.
        }
    \label{fig:pi}
\end{figure}

Each simulation requires a \json-formatted input file
where the parameters are given.
Such files can be created from a master input file,
specifying the parameters for a simulation series.
A master input file can be created from a regular input file
by changing a parameter value into a list of values.
More than one parameter can contain lists,
and all possible combinations of values is found
to create the input files for the simulation series.
To demonstrate,
we use \ilcode{cmsim.json}
in \cref{lst:sim_series_example},
which specifies a simulation series exploring the influence of various parameters.
The ten values for the applied voltage
are specified through a \ilcode{linspace}-command,
while the values for the threshold for breakdown in the channel
and
photoionization (fast mode) enabled
are given in list form.
Furthermore,
\ilcode{simulation_runs} specifies the number of similar simulations,
only differing by \ilcode{random_seed}.
If \ilcode{random_seed} is \ilcode{null},
then each input file is created with a random number as \ilcode{random_seed},
and
when a number is specified,
a range of number is generated,
in this case,
the numbers 1 through 10.
Note that \ilcode{random_seed}
refers to the seed number for initializing the random number generator,
not to the seeds within the \roi.
However, a given \ilcode{random_seed} does corresponds to
a given initial positions of the seed anions.
Defining fixed a \ilcode{random_seed} for each simulation series
makes it easier to see how a change in a given value affects the simulation,
but for analyzing a larger assemble of simulations,
it is usually preferable that the simulations are uncorrelated,
i.e.\ have random initial anion placement.

Individual input files are created by running
the \ilcode{cerman} with the argument \ilcode{ci} (create input)
and specifying which file to expand with \ilcode{-f}:
\begin{lstlisting}[language=fish]
    cerman ci -f <filename>
\end{lstlisting}
This command creates a a number of new files by permutation
of all lists in the given input file.
The permutation of
10 random seeds,
10 applied voltages,
5 breakdown thresholds,
and
2 modes for photoionization
in \cref{lst:sim_series_example}
results in 1000 files.
By default,
the random seed is expanded first,
followed by the needle voltage,
which is is useful to consider when designing a simulation series.
Choosing an appropriate number of values for these two parameters
makes it easier to search for simulation files with given properties.
When expanding the example in \cref{lst:sim_series_example},
the least significant digit \ilcode{??X} indicates random seed number,
the second digit \ilcode{?X?} indicates the needle voltage,
while
the most significant digit \ilcode{X??}
indicates the threshold for breakdown in the streamer channel
and
whether photoionization is enabled or not.

The action \ilcode{pp} (plot parameters)
creates a matrix representation
of the parameter variation in set of input files,
and is used like
\begin{lstlisting}[language=fish]
    cerman pp -g <pattern>
\end{lstlisting}
The argument \ilcode{-g} specifies the pattern to search (or ``glob'') for,
e.g.\ \ilcode{cmsim_?00.json}.
The files are plotted at the $x$-axis and the
varied parameters on the $y$-axis,
see \cref{fig:pi} for an example output.
The name of the output file
is based on the first file in the pattern.

After ensuring that the input parameter values are as desired,
simulations are run using
\begin{lstlisting}[language=fish]
    cerman sims -g <pattern> -m <no>
\end{lstlisting}
which creates a queue of all files matching the pattern,
and simulates a given number in parallel.
For instance, \ilcode{cerman sims -g "cmsim_?5?.json" -m 15},
simulates all input files with the same voltage,
creating a queue where up to 15 separate subprocesses
each run one simulation.
These python processes are single threaded,
and works best if
\ilcode{numpy} is limited to a single thread as well.
Each simulation dumps
the input parameters
and
progress information to a log file.
For each simulation we then have
a parameter file,
a log file,
and
one or more save files.
The files are named by extending the name of the master file,
e.g.\
\begin{lstlisting}[language=fish]
    cmsim.json               # master file
    cmsim_290.json           # input parameters
    cmsim_290.log            # log file
    cmsim_290_gp5.pkl        # save file
    cmsim_290_stat.pkl       # save file
\end{lstlisting}

\subsection{Evaluate results}\label{evaluate-results}

The results are evaluated
by parsing the data stored from one or more simulations.
The input file,
\cref{lst:sim_series_example},
defines two ``save specifications'' as \ilcode{true}, i.e.\ enabled,
each defining various simulation data
to be dumped to disk at given intervals or occasions.
The \ilcode{save_spec} called \ilcode{stat} saves
iteration number,
CPU time,
simulation time,
leading head $z$ position,
number of critical avalanches,
and
the position of each new streamer head,
for every iteration.
This enables evaluation of the shape of the streamer and its propagation speed.
The \ilcode{save_spec} called \ilcode{gp5} saves
most of the data available,
including the position of all the seeds (anions/electrons/avalanches),
for every 5 percent of streamer propagation,
i.e.\ a total of 20 times for a breakdown streamer.
The data is saved using \ilcode{pickle} and can be loaded to
analyze
a given iteration,
a whole simulation,
or by combining data from several simulation.

\textit{Evaluate iterations.}~%
Iteration data can be used to analyze the details of a simulation.
This is particularly useful when evaluating
the validity of the simulation parameters.
Use for instance
\begin{lstlisting}[language=fish]
    cerman pi seedheads -r <start stop> -g <pattern>
\end{lstlisting}
where \ilcode{pi} means ``plot iteration''
and \ilcode{seedheads}
is a scatter plot of avalanches and the streamer head configuration.
The option \ilcode{-r}
controls the range of iterations to plot.
Figure 28 in \cite{Madshaven2018cxjf} shows a number of such plots.

\begin{figure}[b!]
    \centering
    \includegraphics[width=0.45\linewidth]{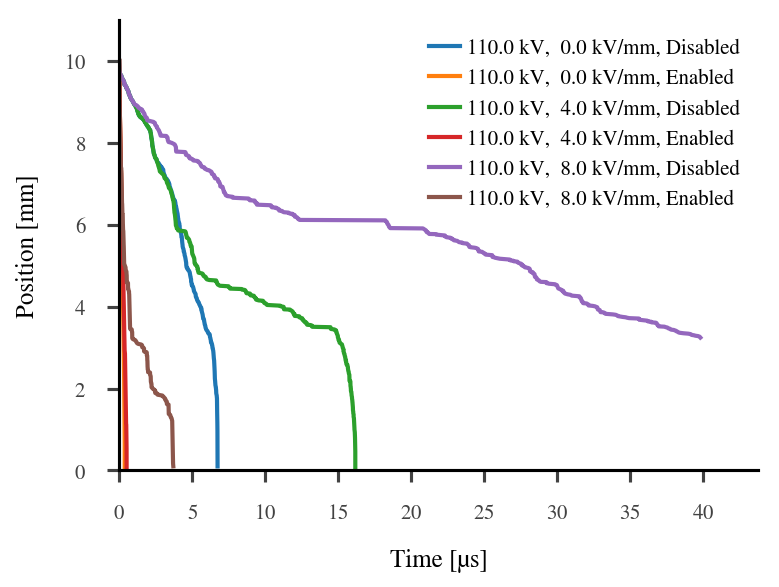}
    \hfill
    \includegraphics[width=0.50\linewidth]{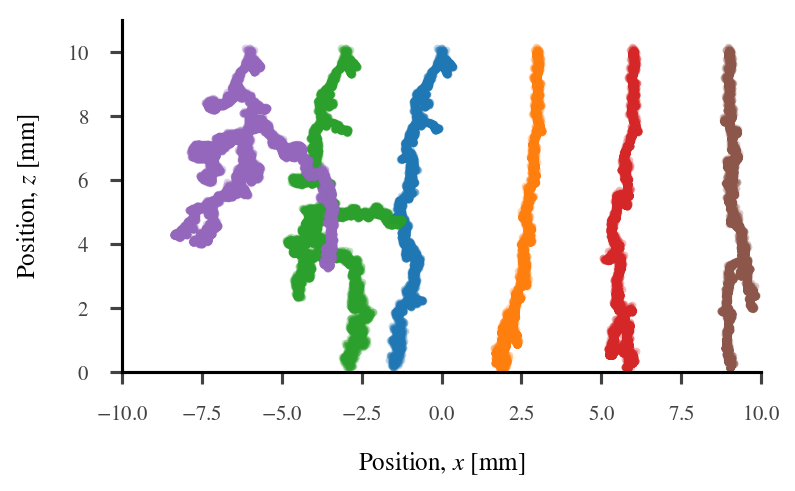}
    \caption{
        (left) Streak plots where ``options'' have been used to limit the $x$-axis
        and to show $V$, $E_'bd'$, and photoionization on the legend.
        (right) Shadow plots where each streamer is plotted
        with an offset and the legend is hidden.
        The legend is the same for both plots.
        }
    \label{fig:streak_trail}
\end{figure}

\begin{figure}
    \centering
    \includegraphics[width=0.49\linewidth]{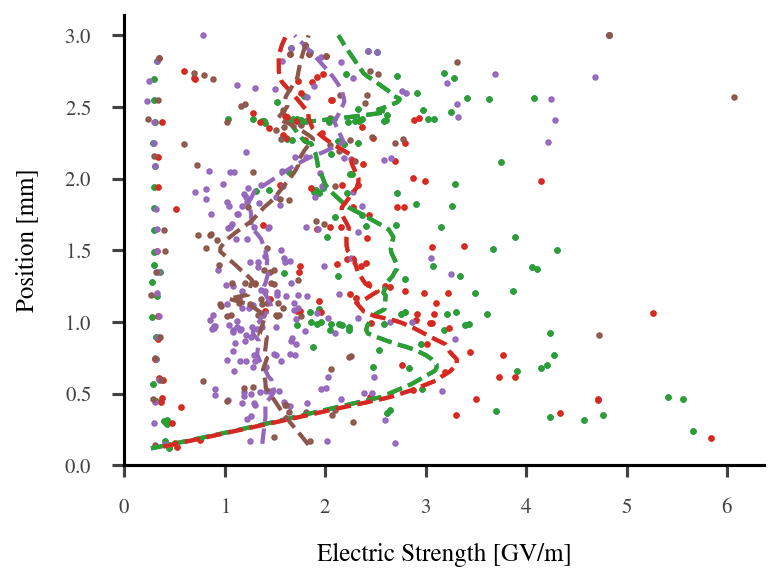}
    \hfill
    \includegraphics[width=0.49\linewidth]{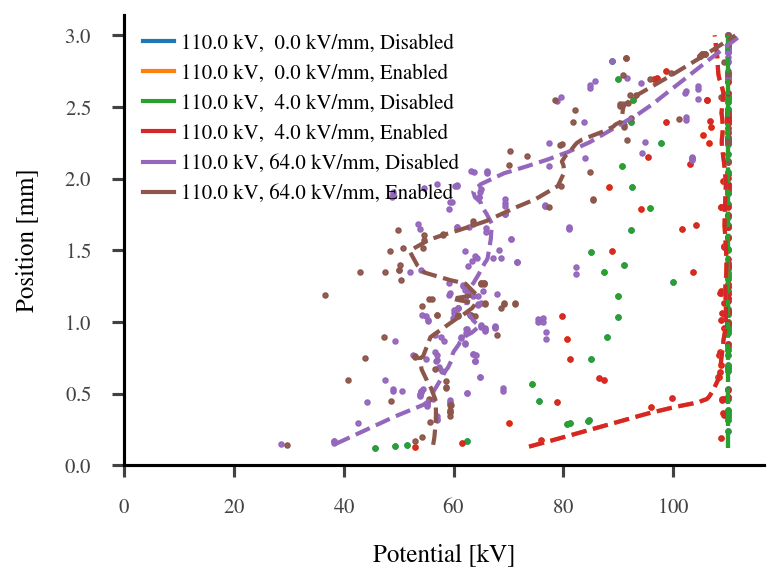}
    \caption{
        The electric strength (left)
        and
        the electric potential (right)
        at the tip of each new streamer head.
        The streamer heads are sampled for every 5~\% of propagation.
        The ``options'' are used to control
        which streamer heads to use for the calculation
        and
        which positions to calculate for.
        }
    \label{fig:scale_estr}
\end{figure}

\textit{Evaluate simulations.}~%
Use \ilcode{ps} for simulation plots.
These are mainly plotted with the $z$-position in the gap on the $y$-axis.
Plotting the $x$- or $y$-position of streamer heads
on the $x$-axis,
using \ilcode{shadow},
gives a ``shadowgraphic'' plot of the streamer:
\begin{lstlisting}[language=fish]
    cerman ps shadow -g <pattern> -o <options>
\end{lstlisting}
Similarly,
plotting the propagation time on the $x$-axis
is done in a \ilcode{streak} plot.
Options can be added to control
the limits/extents of the plot,
the figure size,
the behavior of the legend,
redefining the axis labels,
starting each plot with an offset,
saving the plotted data to a \json-file,
and much more.
Use \ilcode{help} to show available commands and options,
for instance:
\begin{lstlisting}[language=fish]
    cerman help               # for the main script
    cerman ps help            # for plot simulation
    cerman ps shadow -o help  # for shadow plot
\end{lstlisting}

Single simulation data may be of interest,
but it is often better to compare
several simulations in the same plot to visualize
how the input parameters affect the results.
The \ilcode{gp5} save requires a lot of disk space,
but can be very useful in analyzing the data.
For plotting the potential of each head, use
\begin{lstlisting}[language=fish]
    cerman ps headsestr -g <pattern> -o <options>
\end{lstlisting}
The current (active) heads of the streamer is selected,
their potential is scaled
(electrostatic shielding, using a nnls-approach, cf.~\cite{Madshaven2018cxjf}),
and then,
the electric field at their position is calculated.
However, the options can be used to specify the method for scaling
and which positions to calculate the field for,
e.g.\
at the position of each appended (new) streamer head.
The electric field strength
and
electric potential
are presented as scatter plots
against the $z$-position of each given position,
as well as dashed line indicating the average value,
see \cref{fig:scale_estr}.

\begin{figure}
    \centering
    \includegraphics[width=0.8\linewidth]{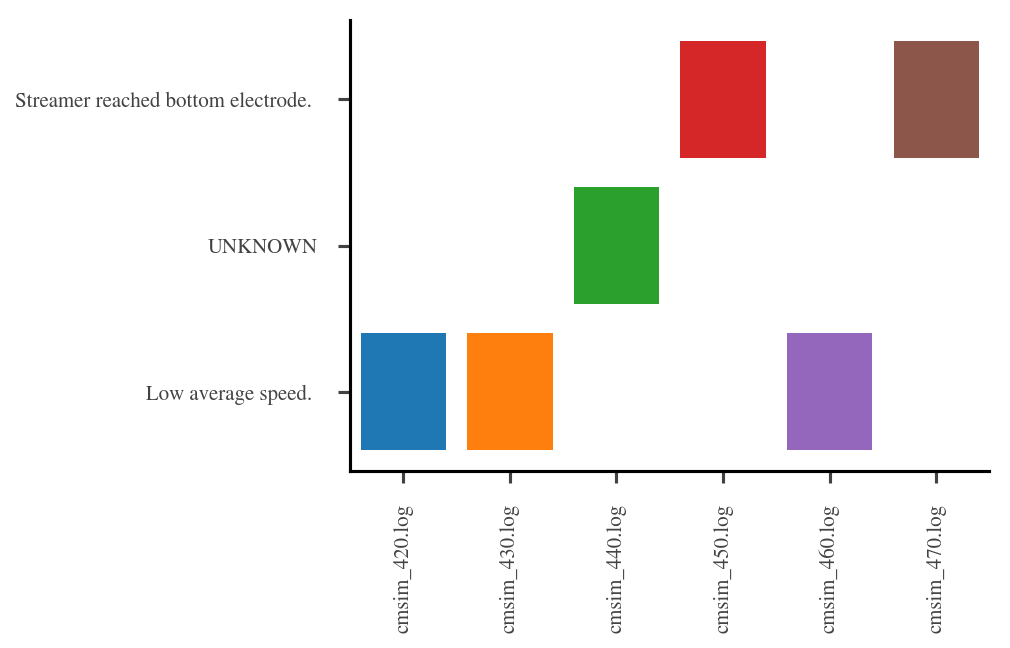}
    \caption{
        Example, parsing the log files to visualize how simulations have terminated.
        ``Unknown'' implies that the simulation is not complete (ongoing or aborted).
        }
    \label{fig:leg_res}
\end{figure}

\begin{figure}
    \centering
    \newcommand{\width}{0.49\linewidth}
    \includegraphics[width=\width]{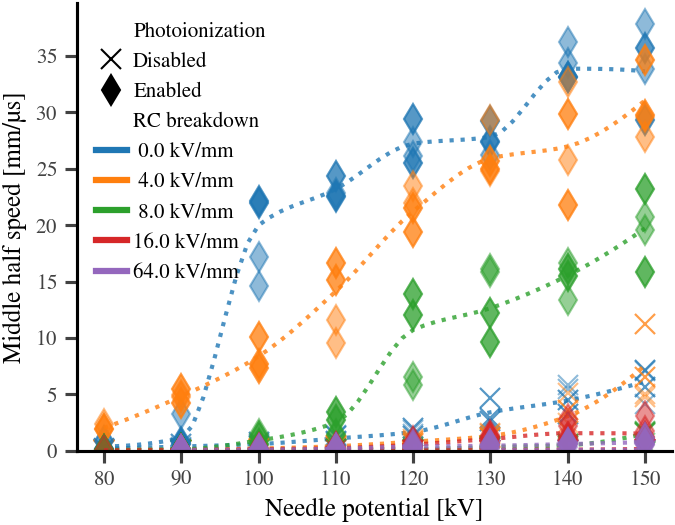}
    \includegraphics[width=\width]{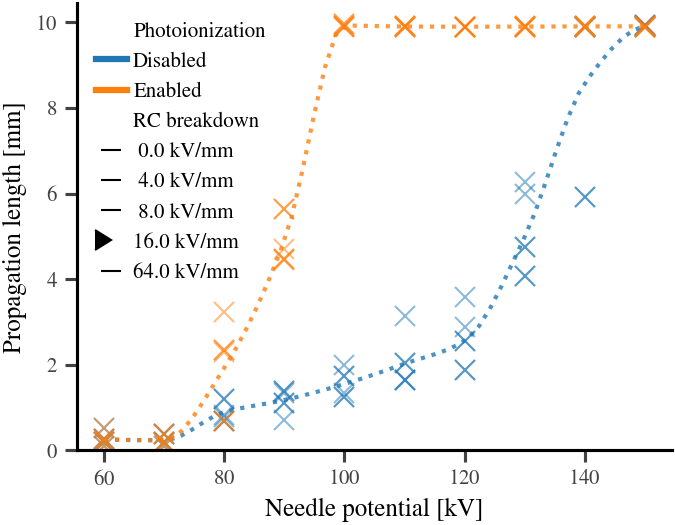}
    \caption{
        Example, combination plots:
        (left) propagation speed using both markers and colors,
        and
        (right) propagation length for only a given value of one parameter
        with the other parameter colored.
        }
    \label{fig:combination}
\end{figure}

\textit{Evaluate a series of simulations.}~%
The difference in the simulated parameters
can be visualized using \ilcode{pp}
and globbing for the \ilcode{pkl}-files or \ilcode{log}-files.
An existing \ilcode{log}-file indicates that a simulation was initiated.
The command
\begin{lstlisting}[language=fish]
    cerman psr -g <pattern>
\end{lstlisting}
parses log files and plots
the reasons why the simulations were terminated.
An example of such a plot is shown in \cref{fig:leg_res}.
This is a good way to verify that
simulations have completed successfully.

When all simulations are completed and verified,
parse all the save files and build a combined database of the results:
\begin{lstlisting}[language=fish]
    cerman ca -g <pattern> -o mode=reload
\end{lstlisting}
The option \ilcode{reload} forces files to be parsed,
even if they already are in the database.
When parsing the save files,
a number of properties are extracted or calculated.
Results
such as
propagation length (ls),
average propagation speed (psa),
inception time (ita),
average jump distance (jda),
simulation time (st),
and
computational time (ct)
are saved in the database.
The results are plotted by using
\begin{lstlisting}[language=fish]
    cerman pr <parameter> -f <file> -o <options>
\end{lstlisting}
The parameter
gives the $x$-axis of the plot,
for instance \ilcode{v} (needle voltage).
The results to plot is given through the option,
e.g.\ \ilcode{-o ykeys=ls_psa}.
The software inspects the database of parsed save files for varied parameters
and
automatically create plots of all possible permutations
using colors and markers.
\Cref{fig:combination} shows a selection of such plots,
where needle voltage is on the $x$-axis
and the other parameters have been added automatically.

As illustrated,
the software has been designed to facilitate
simulating and analyzing
a large simulations series
in a semi-automatic fashion.
The individual simulations have their parameters defined in dedicated files,
while the behavior of the \ilcode{cerman}-script
is defined by command line arguments.

}
%


\section{Model implementation}{\label{sec:implementation}

This section describes the implementation of our model~%
\cite{Madshaven2018cxjf,Madshaven2019c933,Madshaven2020dg8m}
in more detail.
An overview of the model is already given in \cref{sec:avamod},
and \cref{fig:seeds_hyper} is useful for understanding the principal setup,
a needle--plane gap where
electron avalanches grow from single electron seeds.

\textit{The electric field.}~%
The needle electrode and the streamer heads
are modeled as hyperboloids.
The Laplacian electric potential $V_i$ and electric field $\vec{E}_i$
from a streamer head $i$
is calculated analytically in prolate spheroid coordinates~%
\cite{Madshaven2018cxjf}.
For a position $\vec{r}$,
the electric potential $V(\vec{r})$ and field $\vec{E}(\vec{r})$
is given by the superposition principle,
\begin{equation}
    V(\vec{r}) = \sum_i k_i V_i(\vec{r})
    \quad\text{and}\quad
    \vec{E}(\vec{r}) = \sum_i k_i \vec{E}_i(\vec{r})
    \,,
    \label{eq:VrEr}
\end{equation}
where the coefficients $k_i$ are introduced
to account for electrostatic shielding.
An optimization is performed such that
the unshielded potential at the tip of any streamer head $j$
equals the sum of all the shielded potentials at that position,
$V_j(\vec{r_j}) = \sum k_i V_i (\vec{r_j})$~%
\cite{Madshaven2018cxjf}.

\textit{Region of interest (\ROI).}~%
The \roi\ is a cylindrical volume
used to control the position of seeds,
see \cref{fig:seeds_hyper}.
Note, the \emph{seeds} in this respect include
all anions, electrons, and even all avalanches.
The number of seeds in a simulation is given by
the specified density of seeds and the volume of the \roi.
Initially,
the seeds are placed at random positions within the \roi.
When a seed falls behind the \roi,
collides with the streamer,
or creates a critical avalanche,
it is removed and replaced by a new seed.
The new seed is placed with a $z$-position
one \roi-height closer to the plane,
at a random $xy$-position (at a radius less than the \roi\ radius).
The \roi\ volume
is defined by a distance from the $z$-axis,
and a given length above and below the leading streamer head,
which is the part of the streamer closest to the planar electrode.
When a new streamer head is created closer to the plane,
the streamer propagates,
and the \roi\ moves as well.

\textit{Anions, electrons, and electron avalanches.}~%
The electric field $E$ is calculated for each seed
and the seeds are classified as
avalanches ($E \geq E_'c'$),
electrons ($E \geq E_'d'$),
or
anions ($E < E_'d'$),
where
the avalanche threshold $E_'c'$
and
detachment threshold $E_'d'$
are given as input parameters.
Seeds move in the electric field
with a speed dependent on their mobility $\mu$,
which gives the distance the seed moves,
$\Delta \vec{s} = \vec{E} \mu \Delta t$,
for a time step $\Delta t$.
The number of electrons $N_'e'$ in an electron avalanche,
starting from a single electron,
can be calculated by~\cite{Madshaven2018cxjf}
\begin{equation}
    N_'e' = \exp \sum_i \Delta s_i \alpha_i
        =
        \exp  \sum_i E_i \, \mu_'e' \,
        \alpha_'m' \, e^{-{E_\alpha} / {E_i}} \Delta t_i
    \,,
    \label{eq:Ne}
\end{equation}
where
$\alpha$ is the avalanche growth,
$\mu_'e'$ is the electron mobility,
$\Delta t$ is the time step,
and
$i$ is an iteration,
whereas
the avalanche growth parameters
$\alpha_'m'$ and $E_\alpha$
have been obtained from experiments~%
\cite{Madshaven2018cxjf}.
In practice,
however, we only calculate and store the exponent in \cref{eq:Ne},
$Q_'e' = \ln N_'e'$,
\begin{equation}
    Q_'e' = \sum_i \Delta Q_i
      = \sum_i \Delta s_i \alpha_i
    \,,
    \label{eq:Qe}
\end{equation}
for each electron avalanche.
When a low-IP additive is present,
$\alpha$ is modified by adding a factor~\cite{Madshaven2018cxjf}
\begin{equation}
    \alpha_{i,\text{add}} = \alpha_i (1 - x_'add' + x_'add' e^{k_\alpha (I_'b' - I_'a')})
\end{equation}
which is dependent on
the mole fraction of the additive $x_'add'$,
and
the difference in IP between the base liquid $I_'b'$ and the additive $I_'a'$
modified by a factor $k_\alpha$
as prescribed by~\cite{Ingebrigtsen2009fptpt5}.
This is the default setting for the software.
Another model for $\alpha$ given in~\cite{Atrazhev1991d2mg8n} is also implemented:
\begin{equation}
    \alpha_{i,\text{mod}} = \frac{3 e E_\alpha^2}{I_'b' E_i} e^{-\frac{E_\alpha}{E_i}}
    \,.
    \label{eq:alpha_atraz}
\end{equation}
This method is applied in \cref{lst:sim_series_example}.

\textit{Expanding the streamer structure.}~%
According to the Townsend--Meek criterion~\cite{Pedersen1989d2kthj},
streamer breakdown occurs when an avalanche exceeds
a critical number of electrons $N_'c' = \exp(Q_'c')$.
When an avalanche obtains $Q_'e' > Q_'c'$,
we place a new streamer head at its position~%
\cite{Madshaven2018cxjf}.
The initial potential of a new streamer head
is calculated by considering
the capacitance and potential
of the closest existing streamer heads~%
\cite{Madshaven2019c933}.
If adding the new head
implies removing another head (see the paragraph below),
the potential changes slightly,
mimicking transfer of charge.
However,
if both the new and the present head stays,
they share the ``charge'',
which gives a moderate reduction in the potential of both heads.

\textit{Optimizing the streamer structure.}~%
There are three criteria for removing heads.
A streamer head $i$ is removed if
\begin{equation}
    \nu_j(\vec{r}_i) < \nu_j(\vec{r}_j)
    \quad
    \text{or}\quad
    k_i < k_'c'
    \quad
    \text{or}\quad
    \big(
    (\abs{\vec{r}_i - \vec{r}_j} < d_'m')
    \text{ and }
    (z_i > z_j)
    \big)
    \,,
    \label{eq:trimming}
\end{equation}
are satisfied
for any other streamer head $j$~%
\cite{Madshaven2018cxjf}.
The first condition checks whether the tip of one hyperbole ($\vec{r}_i$)
is inside another hyperbole, a collision
(the $\nu$-coordinate describes a hyperboloid, specifically the asymptotic angle).
The second condition removes heads whose potential
are to a high degree shielded by other heads
(if the coefficient $k_i$ in \cref{eq:VrEr} lower than a threshold $k_c$).
The third condition checks whether two streamer heads are closer than $d_'m'$
and should be merged to a single head,
where the one at the highest $z$-coordinate is removed.

\textit{Conduction and breakdown in the streamer channel.}~%
Conduction in the streamer channel
increases the potential of each streamer head $i$ each iteration,
\begin{equation}
    \Delta V_i = V_0 - V_i
    \quad\rightarrow\quad
    V_i = V_0 - \Delta V_i \, e^{- \Delta t / \tau_i}
    \,.
    \label{eq:Vrelax}
\end{equation}
The time constant $\tau_i = RC\tau_0$
(for a given head $i$)
is calculated from
the resistance $R$ in the channel
and
the capacitance $C$ towards the plane%
~\cite{Madshaven2019c933}
\begin{equation}
    R \propto \ell
    \,,\quad\text{and} \quad
    C \propto \left( \ln \frac{4z + 2 r_'s'}{r_'s'} \right)^{-1}
    \,,
    \label{eq:RC}
\end{equation}
where
$\ell$ is the channel length (distance to the needle),
$r_'s'$ is the tip curvature radius of the streamer head,
and
$z$ is the $z$-position of the streamer head.
As such,
each streamer head is treated as an individual RC-circuit,
e.g.\ the three streamer heads in \cref{fig:seeds_hyper}
would each have an individual resistance (channel)
as well as an individual capacitance towards the planar electrode.
If the conduction is low,
the potential difference $\Delta V_i$ increases as the streamer propagates,
and the electric field within the streamer channel
$E_'s' = \Delta V_i / \ell_i$ may increase as well.
If $E_'s'$ exceeds a threshold $E_'b'$,
a breakdown occurs,
equalizing the potential of the streamer head and the needle,
which is achieved by setting $\tau_i = 0$ in \cref{eq:Vrelax} for the given iteration.

\textit{Photoionization.}~%
Photoionization is a possible mechanism
for fast streamer propagation~\cite{Linhjell1994chdqcz}.
We have proposed a mechanism
in which the propagation speed of a streamer increases
if the liquid cannot absorb radiation energy to excited states,
as a result of a strong electric field reducing the ionization potential~%
\cite{Madshaven2020dg8m}.
Since the full model,
considering fluorescent radiation from the streamer head,
and a field-dependent photoionization absorption cross section,
is computationally expensive,
a simpler model is used in the simulations.
Instead we calculate the field strength $E_w$
required
to reduce the ionization potential
below the energy of the fluorescent radiation.
In each iteration,
if the electric field $E$
at the tip of a streamer head $i$ exceeds the parameter $E_w$,
the head is moved
a distance $\Delta \vec{s}_i$
towards the planar electrode,
\begin{equation}
    \Delta \vec{s}_i = - v_w \, \Delta t  \, \Theta \! \left(E(\vec{r}_i) - E_w \right) \, \hat{z}
    \,,
\end{equation}
where
$v_w$ is the photoionization speed,
and
$\Theta$ is the Heaviside step function.
For more details on the entire model, see our previous work~%
\cite{Madshaven2018cxjf,Madshaven2019c933,Madshaven2020dg8m}.

}
%


\section{Current functionality and future prospects}{\label{sec:functionallity}

The main function of the model and the software
is to simulate streamers in a point--plane gap,
using the Townsend--Meek criterion for propagation.
The propagation criterion is met when
electron avalanches obtain a given size.
This model and the algorithm are fixed,
but there are several parameters which can be adjusted.
Changing experiment features such as
needle tip radius, gap size, voltage, liquid properties,
or the parameters of the algorithms,
is straightforward.
Proposals to extend the software
to encompass new functionality
is given in this section.

In~\cite{Madshaven2018cxjf}
we explored the fundamental features off the model,
i.e.\
a streamer consisting of charged hyperbolic streamer heads,
and
streamer growth by electron avalanches initiating from anions.
The model predicts several aspects of streamer propagation,
and shows how they are linked towards the values of given input parameters.
The predicted
propagation speed and the degree of branching
were both lower than expected.
We found how
the speed was dependent
mainly on the number of electron avalanches and their growth,
while the branching was mainly related to
how the streamer heads were configured and managed,
which is mainly controlled
by the parameters $k_'c'$ and $d_'m'$ in \cref{eq:trimming}.

When new streamer heads were added,
their potential was set
assuming a constant electric field within the channel,
resulting in a moderate voltage drop between
the needle electrode and the streamer head%
~\cite{Madshaven2018cxjf}.
To better represent the dynamics of the streamer channel,
an RC-model was developed%
~\cite{Madshaven2019c933}.
In the RC-model,
the potential of new streamer heads is
dependent of the potential of the closest existing streamer heads.
If the conductance of the streamer channel is high,
then the potential of the streamer head is kept close to that of the needle,
giving results comparable to those without the RC-model.
Conversely,
having low conduction regulates the speed of the streamer,
increasing the likelihood that more branches are able to propagate.
Furthermore,
the RC-model also allows for simulation of a breakdown
within the streamer channel itself,
which is likely what occurs during a re-illumination.
This breakdown occurs when the
electric field within the channel exceeds a given threshold.

The importance of photoionization
during a streamer breakdown is unknown.
We explored different aspects of photoionization
in~\cite{Madshaven2020dg8m},
and implemented a model for change to a fast propagating mode.
Molecules excited by energetic processes,
such as electron avalanches,
can relax to a lower energy state by emitting radiation.
We argued that fluorescent radiation can be important,
and modeled how this radiation
can cause ionization in high-field areas,
since the high-field reduces the ionization potential.

In the current implementation,
a square wave voltage is applied to the needle at the beginning of the simulation.
It is easy to change the behavior to a voltage ramp,
from zero to max over a given time.
This can be the basis for a study on streamer inception
where other parameters such as
needle size and the electron properties of the liquid
are investigated as well.
Simulating a dynamic voltage,
such as a lightning impulse,
requires some more work,
but is also achievable.

We have focused on cyclohexane
since many of its properties are well known,
but other non-polar insulating liquids
can be studied by changing relevant parameters.
The seed density $n_'ion'$
is based on
the low-field conductivity $\sigma$
and
electron mobility $\mu_'e'$ of the liquid,
and
the propagation speed scales linearly with both
seed density $n_'ion'$
and
electron mobility $\mu_'e'$.
The electron avalanche growth parameters
are also liquid-dependent,
and
$E_\alpha$ in particular
has a big impact on the results%
~\cite{Madshaven2018cxjf}.
Streamer parameters,
such as
conductivity of the streamer channel
and
streamer head radius,
need to be reevaluated as well for other liquids.
The properties of the streamer channel are also important
to simulate the effects of external pressure,
which mainly affects processes in the gaseous phase~%
\cite{Linhjell2019dvbz}.

The effect of additives with a low ionization potential (IP) are modeled
as causing an increase in electron avalanche growth%
~\cite{Ingebrigtsen2009fptpt5,Madshaven2018cxjf}.
Other additives can easily be used
as long as the IP of both the base liquid and the additive are known.
Low-IP additives are known to facilitate
the propagation of slow streamers
and
to increase the acceleration voltage,
possibly as a result of increased branching~%
\cite{Lesaint2000c4xf84},
however the mechanisms involved are not known.
It is possible that low-IP additives
are sources of electrons that can initiate avalanches,
produced
for instance through photoionization
or fluctuations in the electric field.
Such mechanisms can be added to the model and simulated,
but will require some work.
Furthermore,
the mechanisms of added electron scavengers
can also be interesting for further investigation,
and particularly if negative streamers are to be simulated~%
\cite{Ingebrigtsen2009fptpt5}.

Our primary concern have been with positive streamers,
since these are more likely to lead to a complete breakdown
than negative streamers.
The model relies on electrons detaching from anions,
moving towards regions where the field is higher,
and then forming electron avalanches.
The polarity of our model can easily be reversed,
however,
the electrons would then drift away
and be unable to form avalanches.
As such,
a model for creation of new electrons is needed
to simulate negative streamers with the software.
Charge injection from the needle and the streamer
can be one such mechanism~\cite{Smalo2011dmkszm}.
Another option is to model
electron generation (charge separation)
in the high-field region
surrounding the needle and the streamer.
Such mechanisms are interesting
for simulating positive streamers as well.

The hyperbole approximation simplifies
the calculation of the electric field,
both from the needle electrode and from the streamer heads.
Other experimental geometries,
such as plane--needle--plane,
or even more realistic real-life geometries
can be implemented.
The challenge is
to set the correct shielding or scaling of the streamer heads
according to the influence of the geometry.
Simpler geometric restrictions are easier to implement,
for instance by manipulating the \roi.
The streamer can be restricted to a tube
by setting a low value for the maximum radius of the \roi.
Another method is
setting the ``merge threshold'' very high,
such that the streamer is restricted to a single channel with a single head,
which can be representative for a streamer in a tube~%
\cite{Massala1998cfk7km}.

There are many mechanisms that
can be added to investigate different
methods of streamer propagation,
for instance effects of
Joule heating or electro-hydrodynamic cavitation~%
\cite{Lesaint2018gfhkt8}.
There are also several parts of the existing model that can be improved.
Better calculation and balancing of charges and energy
would greatly improve the model.
For instance
an electric network model
where the streamer is consisting of several interconnected parts,
in contrast to the current implementation
where all the streamer heads are individually ``connected'' to the needle.
Such an approach can give a better understanding
of the charge flow in the different parts and branches of the streamer,
as well as a better representation of the electric field.
Development towards a model for
a space-charge limited field~\cite{Boggs2005dmrb95}
can further improve the electric field representation,
however,
possibly at a high computational cost.

}
%


\section{Summary}{\label{sec:conclusion}

We present a software
for simulating the propagation of positive streamers
in a needle--plane gap
insulated by a dielectric liquid.
The model is based on the Townsend--Meek criterion
in which an electron avalanche
have to obtain a given size for the streamer to propagate.
The software was developed and used for simulating
our models for
electron avalanche growth~\cite{Madshaven2018cxjf},
conductance and capacitance in the streamer channels~\cite{Madshaven2019c933},
as well as
photoionization in front of the streamer~\cite{Madshaven2020dg8m}.
From the examples on how to set up, run, and evaluate simulations,
others can recreate our previous results
or create their own set of simulations.
Furthermore,
the overview of the implementation and algorithm
serves as a good starting point for others to change or extend the
functionality of the software.

}
%


\section*{Acknowledgment}{
This work has been supported by
The Research Council of Norway (RCN),
ABB and Statnett,
under the RCN contract 228850.
}

\appendix

\section{The algorithm}{\label{sec:algorithm}

\begin{figure}
  \centering
  \includegraphics[width=\linewidth]{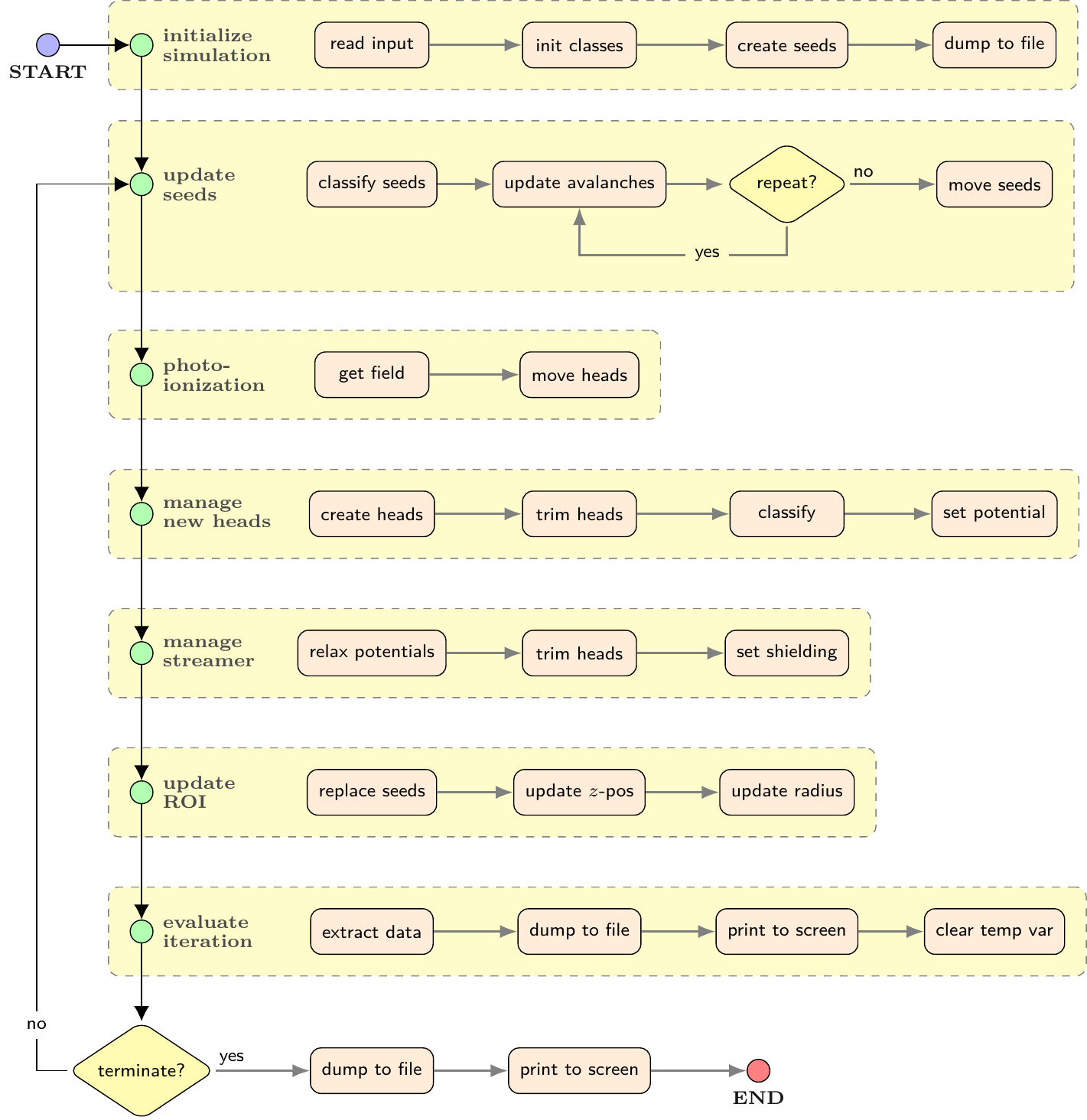}
  \caption{
    The simulation algorithm consists of initialization,
    a loop where the seeds and the streamer structure is updated,
    and then a finalization.
    It the loop,
    first the seeds (anions, electrons avalanches) are affected by the electric field,
    then the streamer structure is modified and this changes the electric field,
    finally the region of interest (\roi) is updated
    and the data from the iteration is evaluated.
    The loop concludes when one (of several) criterion is met,
    typically low propagation speed or reaching the opposing electrode.
    Details on each step is found in \protect\cref{sec:algorithm}.
      }
    \label{fig:alg_main_new}
\end{figure}

This section describes the algorithm used to implement the model
in more detail,
essentially
each part of \cref{fig:alg_main_new},
while referencing relevant parts from \cref{sec:implementation}.

\textit{Initialize simulation.}~%
The simulation input parameters are read and used to initialize classes for
code profiling,
simulation logging,
calculation of avalanche growth,
the needle,
the streamer,
the region of interest (\ROI),
the seeds (anions, electron, avalanches),
how to save data,
and
how to evaluate simulation data.
The initiation of the log file includes dumping
the input parameters to the file.
Given the \ROI\ volume and the seed density,
a number of seeds is created and placed within the \ROI\ at random positions.
Then, the save files are initialized by dumping the initial data,
mainly information concerning the needle and the seeds.

\textit{Update seeds.}~%
The electric field $E$ is calculated for each seed (applying \cref{eq:VrEr})
and the seeds are classified as avalanches, electrons, and anions.
All the avalanches move in the electric field and grow in size (see \cref{eq:Ne}).
The procedure is repeated until
an avalanche collides with a streamer head,
an avalanche meets the Townsend--Meek criterion,
or a total of $N_"msn"$ repetitions has been performed.
The ``time'' spent in this inner loop sets the time step
for the current iteration of the main simulation loop.
Finally,
all the electrons and anions are moved.
The inner loop over just the avalanches save significant computational time
since
the calculation of the electric field is the most expensive part of the computation
and
the avalanches is usually a small fraction of all the seeds.

\textit{Photoionization.}~%
The electric field at the tip of each streamer head is calculated
and compared with the threshold for photoionization.
Each streamer head where $E > E_w$ is ``moved''
a distance $v_w \Delta t$ towards $z = 0$.
Moving implies creating a new head,
setting the potential by ``transferring charge'',
and removing the old head.

\textit{Manage new heads.}~%
For each critical avalanche a new head is created.
If the simulation time step is set sufficiently low,
there is usually zero or just one new head.
The new head is discarded if it satisfies
any of the criteria in \cref{eq:trimming},
however,
if adding it will cause another to be removed (later),
the new head is classified as ``merging''.
If none of the criteria in \cref{eq:trimming}
is met by adding the new head,
i.e.\ all heads are kept,
then it is a ``branching'' head,
since adding it is potentially the start of a new branch.
The potential of ``merging'' is set by ``charge transfer''
from the closest head,
while ``branching'' heads have their potential set by
``sharing charge'' with the closest head,
where the latter method also modifies the potential of the existing head~%
\cite{Madshaven2019c933}.

\textit{Manage streamer.}~%
The potential of each streamer head is relaxed
towards the potential of the needle
by applying \cref{eq:Vrelax}.
This step increases the potential of the streamer head
to the potential of the needle
when a breakdown in the channel occurs.
As mentioned above,
the calculation of the electric field for the seeds is computationally expensive,
and it actually scales with both the number of seeds and the number of streamer heads.
It is therefore preferable to keep the number of heads to a minimum.
Superfluous heads are trimmed according to the criteria in \cref{eq:trimming}.
Then,
the electrostatic shielding is set for the trimmed structure.

\textit{Update \roi.}~%
Seeds that have
moved behind the \roi,
collided with the streamer,
or lead to a critical avalanche
are removed and replaced by a new seed.
When a seed is replaced,
the new seed is placed a distance,
equal to the height of the \roi,
closer to the plane,
at a random $xy$-position
within the \roi\ radius.
If the streamer has moved closer to the plane,
the \roi\ moves as well,
and seeds behind the new position are replaced.
If a streamer head is close to the edge of the \roi,
the \roi\ expands towards the maximum radius.
New seeds are created at random position within the expanded region.

\textit{Evaluate iteration.}~%
Iteration data is extracted from the various classes to be saved for later use.
The data is used to
evaluate whether any stop condition is fulfilled,
and stored to dedicated classes.
Which data to store and how often to sample the data is controlled by the user input.
The saved data is dumped to a file at regular intervals
to keep memory requirements of the program low.
Information to monitor the progress is printed to the screen, at regular intervals.
Finally, a number of temporary variables, relevant only to the iteration is cleared,
and the program is prepared for a new iteration.

\textit{Terminate?}~%
If none of the conditions for stopping a simulation are met,
the next iteration is performed.
These conditions includes
low streamer speed,
streamer head close to the plane (breakdown),
simulation time exceeded,
computation time exceeded,
and more.
When a criterion is met,
any unsaved data is dumped to disk,
and a final logging to file and screen is performed,
before the program terminates.

}
%


\section{Parameters}{\label{sec:parameters}

\begin{table}
    \centering
    \caption{
        Main parameters for simulation program.
        The \emph{experimental conditions} specifies
        an overvoltage applied to  medium size needle--plane gap.
        The values of the physical parameters in
        \emph{seeds and avalanches}
        and
        \emph{streamer structure}
        are justified in our previous work~%
        \protect\cite{Madshaven2018cxjf,Madshaven2019c933,Madshaven2020dg8m}.
        Parameter values related to the
        \emph{simulation algorithm},
        or the model in general,
        have also been discussed in previous work.
        See further description in \protect\cref{sec:parameters}.
        }


\parbox{1.4\linewidth}{%
\renewcommand{\arraystretch}{0.75}
\hspace*{-10 ex}  
\begin{tabularx}{0.9\linewidth}{
    >{\raggedright\small}X
    >{\ttfamily\small}X
    >{\centering\arraybackslash\hsize=.15\hsize\small}X
    >{\small\centering\arraybackslash\hsize=.40\hsize}X
    }%
    \midrule
        Property
    &   Keyword
    &   \makebox[0pt][c]{Symbol}
    &   Default
    \\
    \midrule   
    \vspace*{-0.3 ex}\textit{Experimental conditions}&&&\\
        Distance from needle to plane
    &   gap_size
    &   $d_'g'$
    &   \SI{10}{\milli\metre}
    \\
        Voltage applied to needle
    &   needle_voltage
    &   $V_'n'$
    &   \SI{100}{\kilo\volt}
    \\
        Needle tip radius
    &   needle_radius
    &   $r_'n'$
    &   \SI{6.0}{\micro\metre}
    \\
    \vspace*{-0.3 ex}\textit{Seeds and avalanches}&&&\\
        Seed number density
    &   seeds_cn
    &   $n_'seeds'$
    &   \SI[per-mode=reciprocal]{2e12}{\per\metre^{3}}
    \\
        Anion mobility
    &   liquid_mu_ion
    &   $\mu_'ion'$
    &   \SI{0.30}{\milli\metre^{2}\per{\volt\second}}
    \\
        Electron mobility
    &   liquid_mu_e
    &   $\mu_'e'$
    &   \SI{45}{\milli\metre^{2}\per{\volt\second}}
    \\
        Liquid low-field conductivity
    &   liquid_sigma
    &   $\sigma_'ion'$
    &   \SI{0.20}{\pico\siemens\per\metre}
    \\
        Electron detachment threshold
    &   liquid_Ed_ion
    &   $E_'d'$
    &   \SI{1.0}{\mega\volt\per\metre}
    \\
        Growth calculation method
    &   alphakind
    &   --
    &   \texttt{I2009}
    \\
        Critical avalanche threshold
    &   Q_crit
    &   $Q_'c'$
    &   \SI{23}{ }
    \\
        Electron multiplication threshold
    &   liquid_Ec_ava
    &   $E_'c'$
    &   \SI{0.2}{\giga\volt\per\metre}
    \\
        Electron scattering constant
    &   liquid_Ealpha
    &   $E_\alpha$
    &   \SI{1.9}{\giga\volt\per\metre}
    \\
        Max avalanche growth
    &   liquid_alphamax
    &   $\alpha_'m'$
    &   \SI[per-mode=reciprocal]{130}{\per\micro\metre}
    \\
        Additive IP diff. factor
    &   liquid_k1
    &   $k_\alpha$
    &   \SI{2.8}{\eV^{-1}}
    \\
        Base liquid IP
    &   liquid_IP
    &   $I_'b'$
    &   \SI{10.2}{\eV}
    \\
        Additive IP
    &   additive_IP
    &   $I_'a'$
    &   \SI{7.1}{\eV}
    \\
        Additive mole fraction
    &   additive_cn
    &   $x_'add'$
    &   \SI{0.00}{}
    \\
    \vspace*{-0.3 ex}\textit{Streamer structure}&&&\\
        Streamer head tip radius
    &   streamer_head_radius
    &   $r_'s'$
    &   \SI{6.0}{\micro\metre}
    \\
        Minimum field in streamer channel
    &   streamer_U_grad
    &   $E_'s'$
    &   \SI{2.0}{\kilo\volt\per\milli\metre}
    \\
        Streamer head merge distance
    &   streamer_d_merge
    &   $d_'m'$
    &   \SI{25}{\micro\metre}
    \\
        Electrostatic shielding threshold
    &   streamer_scale_tvl
    &   $k_'c'$
    &   \SI{0.20}{}
    \\
        Photoionization threshold field
    &   streamer_photo_efield
    &   $E_w$
    &   \SI{3.1}{GV/m}
    \\
        Photoionization added speed
    &   streamer_photo_speed
    &   $v_w$
    &   \SI{20}{km/s}
    \\
        Data type for field calculation
    &   efield_dtype
    &   --
    &   \texttt{float32}
    \\
        RC-model time constant
    &   rc_tau0
    &   $\tau_0$
    &   \SI{1}{\micro s}
    \\
        RC resistance model
    &   rc_resistance
    &   --
    &   \texttt{linear}
    \\
        RC capacitance model
    &   rc_capacitance
    &   --
    &   \texttt{hyperbole}
    \\
        RC breakdown field
    &   rc_breakdown
    &   $E_'bd'$
    &   \SI{6}{kV/mm}
    \\
    \vspace*{-0.3 ex}\textit{Simulation algorithm}&&&\\
        Time step of avalanche movement
    &   time_step
    &   $\Delta t$
    &   \SI{1.0}{\pico\second}
    \\
        Max avalanche steps per iteration
    &   micro_step_no
    &   $N_"msn"$
    &   \SI{100}{}
    \\
        Seed for random number generator
    &   random_seed
    &   --
    &   \texttt{None}
    \\
        Number of similar simulations
    &   simulation_runs
    &   --
    &   10
    \\
        ROI -- behind leading head
    &   roi_dz_above
    &   $z^{+}_"roi"$
    &   \SI{1.0}{\milli\metre}
    \\
        ROI -- in front of leading head
    &   roi_dz_below
    &   $z^{-}_"roi"$
    &   \SI{1.0}{\milli\metre}
    \\
        ROI -- radius from center
    &   roi_r_max
    &   $r_"roi"$
    &   \SI{3.0}{\milli\metre}
    \\
        Stop -- low streamer speed
    &   stop_speed_avg
    &   $v_'min'$
    &   \SI{100}{\metre\per\second}
    \\
        Stop -- streamer close to plane
    &   stop_z_min
    &   $z_'min'$
    &   \SI{50}{\micro\metre}
    \\
        Stop -- avalanche time
    &   stop_time_since_avalanche
    &   $t^'ava'_'max'$
    &   \SI{100}{\nano\second}
    \\
      Sequential start number
    & seq_start_no
    & --
    & 0
    \\
      Enable profiling of code
    & profiler_enabled
    & --
    & False
    \\
      Interval -- dump save data to file
    & file_dump_interv
    & --
    & 500
    \\
      Interval -- display data on screen
    & display_interv
    & --
    & 500
    \\
      Level of logging to file
    & log_level_file
    & --
    & 20
    \\
      Level of logging to console
    & log_level_console
    & --
    & 20
    \\
    \midrule%
\end{tabularx}%
\hspace*{-10 ex}
}%

    \label{tab:parameters}
\end{table}

The parameters for a simulation is supplied by the user
in a \json-formatted file (see \cref{lst:sim_series_example}).
A list of all important parameters are shown in \cref{tab:parameters}.
The potential, position and size of the needle
are important parameters in an experiment
and included in the first section,
followed by parameters controlling
the creation and movement of seeds (anions, electrons, and avalanches).
The third section contains parameters related to
the threshold for electron detachment,
electron avalanche growth,
and
critical avalanche size.
The parameters for the streamer can be split in two groups.
The first group controls creation of the streamer heads
and how the streamer heads are treated in relation to each other,
whereas the second groups is related to the RC-model,
controlling the potential of the streamer heads
and the electric field in the streamer channel.
There is also the option to choose whether
the electric field from the streamer,
acting on the seeds,
is calculated using 32- or 64-bit precision.
The latter requires about twice the time to compute.
The parameters in the last section
are mainly related to the simulation algorithm itself.
The time step and the number of steps per loop are essential for a good results.
The random seed, used to initialize the random number engine,
controls the initial placement of seed anions.
Choosing the same random number enables the study of,
for instance, changing voltage with the same initial seed configuration.
Conversely,
not setting the random number
makes the simulations uncorrelated,
which is better for statistics.
The size of the ROI decides how many seeds that are included in a simulation.
For instance,
an ROI of \SI{10}{mm^3} in combination with a seed density of \SI{2e12}{\per m^3}
results in 20~000 seeds generated,
a size which is easily treated by most computers.
Finally,
several parameters can be used to control how long a simulations will run,
or to stop a simulation if the streamer stops or reaches the other electrode.
There are also parameters controlling
how often information is logged,
and how detailed the logging should be.

}
%


\section{Example files}{\label{sec:examplefiles}

This appendix contains a number of examples of possible simulations.
The files in \cref{lst:baseline,lst:naidis,lst:rc,lst:pi}
are all included in the folder \ilcode{examples} on GitHub~\cite{streamergit}.

\vspace*{1 ex}
\indent{\bfseries \Cref{lst:baseline}:}
    The example file \ilcode{small_gap.json} specifies
    a small gap and a range of voltages
    along with
    many parameter values equal to the defaults
    (cf.\ \cref{tab:parameters}).
    Although all values used in a simulation is stored in the log,
    it is nice to be explicit in the input as well.
    By specifying 10 \ilcode{simulation_runs} and 10 voltages,
    a total of 100 simulations is created from this file.
    Each simulation initiated with the seeds at uncorrelated positions
    since \ilcode{random_seed} is \ilcode{none}.
    \\
\indent{\bfseries \Cref{lst:naidis}:}
    The example file \ilcode{small_gap_mod.json}
    builds on \cref{lst:baseline},
    but a number of parameters are modified,
    notably the seed density and the electron avalanche parameters,
    as well as several parameters for the streamer.
    All data is stored every
    5th percent of propagation by \ilcode{gp5}
    and
    every 100th nanosecond by \ilcode{ta07}.
    Storing the properties of tens of thousands of seeds
    enables plotting of the development of seeds,
    but also requires a lot of disk space.
    Specifying \ilcode{streamer} saves all the streamer heads
    every 0.1 percent of propagation
    and is used to evaluate the development of the streamer.
    Since \ilcode{random_seed} is 1
    and \ilcode{simulation_runs} is 10,
    a range of \ilcode{random_seed} from 1 through 10 is generated,
    giving
    the same 10 initial seed positions for each voltage.
    \\
\indent{\bfseries \Cref{lst:rc}:}
    The example file \ilcode{rc.json}
    specifies the low and high conductivity within the streamer channel,
    and
    a range of threshold fields for electric breakdown in the streamer channel.
    The \ilcode{linspace}-command is a convenient method for
    creating a list of values.
    By combining 2 values for the conductivity and 5 values for breakdown
    with 10 values for the voltage,
    a total of 100 simulations are created from this file.
    \\
\indent{\bfseries \Cref{lst:pi}:}
    The example file \ilcode{pi.json} specifies
    electrical breakdown in the streamer channel,
    with and without photoionization enabled,
    The keyword \ilcode{user_comment}
    does not affect the simulations,
    but can be convenient to set.
    600 input files are created when expanding this file
    (100 voltages,
    with and without photoionization,
    for three different breakdown thresholds).
    \\

\begin{lstfloat}[ht!]
    \caption{
        The example file \protect\ilcode{small_gap.json} specifies
        simulations similar to the baseline studies
        in section~3.1 in~\protect\cite{Madshaven2018cxjf}.
        }
    \vspace*{1.5 ex}
\begin{lstlisting}[language=json]
{   "gap_size": 0.003,
    "needle_radius": 6e-06,
    "needle_voltage": [30e3, 40e3, 50e3, 60e3, 70e3,
                       80e3, 90e3, 100e3, 110e3, 120e3],
    "liquid_name": "cyclohexane",
    "liquid_alphamax": 200e06,
    "liquid_Ealpha": 3.0e09,
    "seeds_cn": 2e12,
    "Q_crit": 23,
    "streamer_head_radius": 6e-06,
    "streamer_U_grad": 2e+06,
    "streamer_d_merge": 50e-6,
    "streamer_scale_tvl": 0.2,
    "streamer_photo_enabled": false,
    "random_seed": null,
    "simulation_runs": 10,
    "stop_sim_time": 100e-06,
    "stop_time_since_avalanche": 1e-07,
    "stop_speed_avg": 100}
\end{lstlisting}
    \label{lst:baseline}
\end{lstfloat}

\begin{lstfloat}[hb!]
    \caption{
        The example file \protect\ilcode{small_gap_mod.json} specifies
        simulations similar to
        the attempts to facilitate branching
        in section~3.5 in~\protect\cite{Madshaven2018cxjf}.
        }
    \vspace*{1.5 ex}
\begin{lstlisting}[language=json]
{   "gap_size": 0.003,
    "needle_radius": 6e-06,
    "needle_voltage": "linspace(30e3, 120e3, 10)",
    "liquid_name": "cyclohexane",
    "liquid_alphamax": 130e06,
    "liquid_Ealpha": 1.9e09,
    "seeds_cn": 8e12,
    "streamer_head_radius": 3e-06,
    "streamer_U_grad": 8e+06,
    "streamer_d_merge": 12.5e-6,
    "streamer_scale_tvl": 0.1,
    "streamer_photo_enabled": false,
    "random_seed": 1,
    "simulation_runs": 10,
    "save_specs_enabled": {
        "stat": true,
        "streamer": true,
        "gp5": true,
        "ta07": true}}
\end{lstlisting}
    \label{lst:naidis}
\end{lstfloat}

\begin{lstfloat}[ht!]
    \caption{
        The example file \protect\ilcode{rc.json} specifies
        simulations similar to those performed
        in section~4 in~\protect\cite{Madshaven2019c933}.
        }
    \vspace*{1.5 ex}
\begin{lstlisting}[language=json]
{   "gap_size": 0.003,
    "needle_radius": 6e-06,
    "needle_voltage": "linspace(30e3, 120e3, 10)",
    "liquid_name": "cyclohexane",
    "liquid_alphamax": 130e06,
    "liquid_Ealpha": 1.9e09,
    "streamer_head_radius": 6e-06,
    "streamer_U_grad": 0,
    "streamer_d_merge": 50e-06,
    "streamer_scale_tvl": 0.1,
    "rc_tau0": [1e-8, 1e-4],
    "rc_resistance": "linear",
    "rc_capacitance": "hyperbole",
    "rc_breakdown": [0, 4e6, 8e6, 16e6, 64e6],
    "random_seed": null,
    "save_specs_enabled": {
        "streamer": true,
        "stat": false}}
\end{lstlisting}
    \label{lst:rc}
\end{lstfloat}

\begin{lstfloat}[hb!]
    \caption{
        The example file \protect\ilcode{pi.json} specifies
        simulations similar to those performed
        in section~7 in~\protect\cite{Madshaven2020dg8m}.
        }
    \vspace*{1.5 ex}
\begin{lstlisting}[language=json]
{   "user_comment": "Photoionization and breakdown",
    "gap_size": 0.010,
    "needle_radius": 6e-06,
    "needle_voltage": "linspace(60e3, 150e3, 100)",
    "liquid_alphamax": 130e06,
    "liquid_Ealpha": 1.9e09,
    "seeds_cn": 2e+12,
    "streamer_U_grad": 0,
    "streamer_d_merge": 50e-06,
    "streamer_scale_tvl": 0.1,
    "streamer_photo_enabled": [false, true],
    "streamer_photo_efield": 3.1e9,
    "streamer_photo_speed": 20e3,
    "rc_tau0": 1e-6,
    "rc_breakdown": [0e6, 4e6, 64e6],
    "simulation_runs": 1,
    "random_seed": null,
    "file_dump_interv": 500,
    "display_interv": 500,
    "save_specs_enabled": {
        "streamer": true,
        "stat": false}}
\end{lstlisting}
    \label{lst:pi}
\end{lstfloat}

}
%

{
\clearpage

}

\end{document}